\documentclass{aastex631}

\begin{document}

\title{NICER observes the full Z-track in GX 13+1}

\author{Mohamad Ali Kaddouh}
\affiliation{Department of Physics \& Astronomy, Wayne State University, 666 West Hancock Street, Detroit, MI 48201, USA}
\author[0000-0003-0440-7978]{Malu Sudha}
\affiliation{Department of Physics \& Astronomy, Wayne State University, 666 West Hancock Street, Detroit, MI 48201, USA}
\author[0000-0002-8961-939X]{Renee M.\ Ludlam}
\affiliation{Department of Physics \& Astronomy, Wayne State University, 666 West Hancock Street, Detroit, MI 48201, USA}

\begin{abstract}

We present the temporal analysis of the persistent neutron star low-mass X-ray binary (NS LMXB) GX 13+1 using NICER data. Classification of this source has been ambiguous so far. 
We investigate the evolution of the source in its hardness-intensity diagram (HID) and power density spectra (PDS) of the 0.5--10 keV NICER archival data. For the first time, we detect the source tracing out the entire Z-track, distinctly identifying the horizontal branch (HB), normal branch (NB) and flaring branch (FB). We also detect a peaked noise component in the PDS at $\sim$ 5.4 Hz, which appears to be present when the source is either in the NB or FB. We note a positive slope of the HB in the HID which could be due to either the high intrinsic absorption of the source or the stronger contribution of the soft spectral components in the soft energy domain.

\end{abstract}


\section{Introduction} \label{sec:intro}

NS LMXBs are composed of a neutron star accreting matter from a companion star that has mass $\lesssim 1$ M$_\odot$. They are classified into Z (X-ray luminosity $\sim$ 0.5--1 L$_{Edd}$) and atoll (X-ray luminosity $\sim$ 0.5--1 L$_{Edd}$) sources based on the shape these sources trace out in the HID/color-color diagram (CCD) \citep{hasinger1989,done2007}. Z sources have three branches - HB, NB and FB with each branch exhibiting characteristic quasi-periodic oscillations (QPOs), very low frequency noise (VLFN) and band limited noise (BLN)/low frequency noise (LFN) components \citep{van2006}. Atoll sources have mainly two branches - the island state and the banana branch. Similar to Z sources, each branch exhibits characteristic oscillations and noise components. 

The classification of the NS LMXB GX 13+1 has been ambiguous so far. The track traced out by the source on the HID/CCD has been incomplete, thus a clear identification has yet to be made. Historically, the source has been analyzed under the assumption that it is either an atoll \citep{hasinger1989, schnerr2003} or Z source \citep{homan1998, homan2004, fridriksson2015} having shown timing features that could be interpreted under each class. Our archival analysis of NICER data in the 0.5--10 keV energy bands reveals the entire Z-track for GX~13+1.


\section{Observation and Data Reduction} \label{sec:obs_data}
We have utilized the NICER archival observations of GX 13+1, starting from 2023, Feb 23 to 2024, April 30 for a total of 22 days. Data were reduced using the `nicerl2' routine and events that occurred when the particle background was low were selected for further analysis. Barycentric correction was performed on the event files using `barycorr'. STINGRAY was used for performing the timing analysis \citep{bachetti2021, huppenkothen2019a,huppenkothen2019b}. HIDs were obtained using the 2--3.8 keV, 3.8--6.8 keV, and 2--6.8 keV energy bands with a binsize of 64 s. 
Averaged PDS was generated using 8 s segments with a time bin size of 1/256 s resulting in a Nyquist frequency of 128 Hz and a frequency resolution of 0.125 Hz.
\begin{figure}
    \centering
    \includegraphics[width=0.8\linewidth]{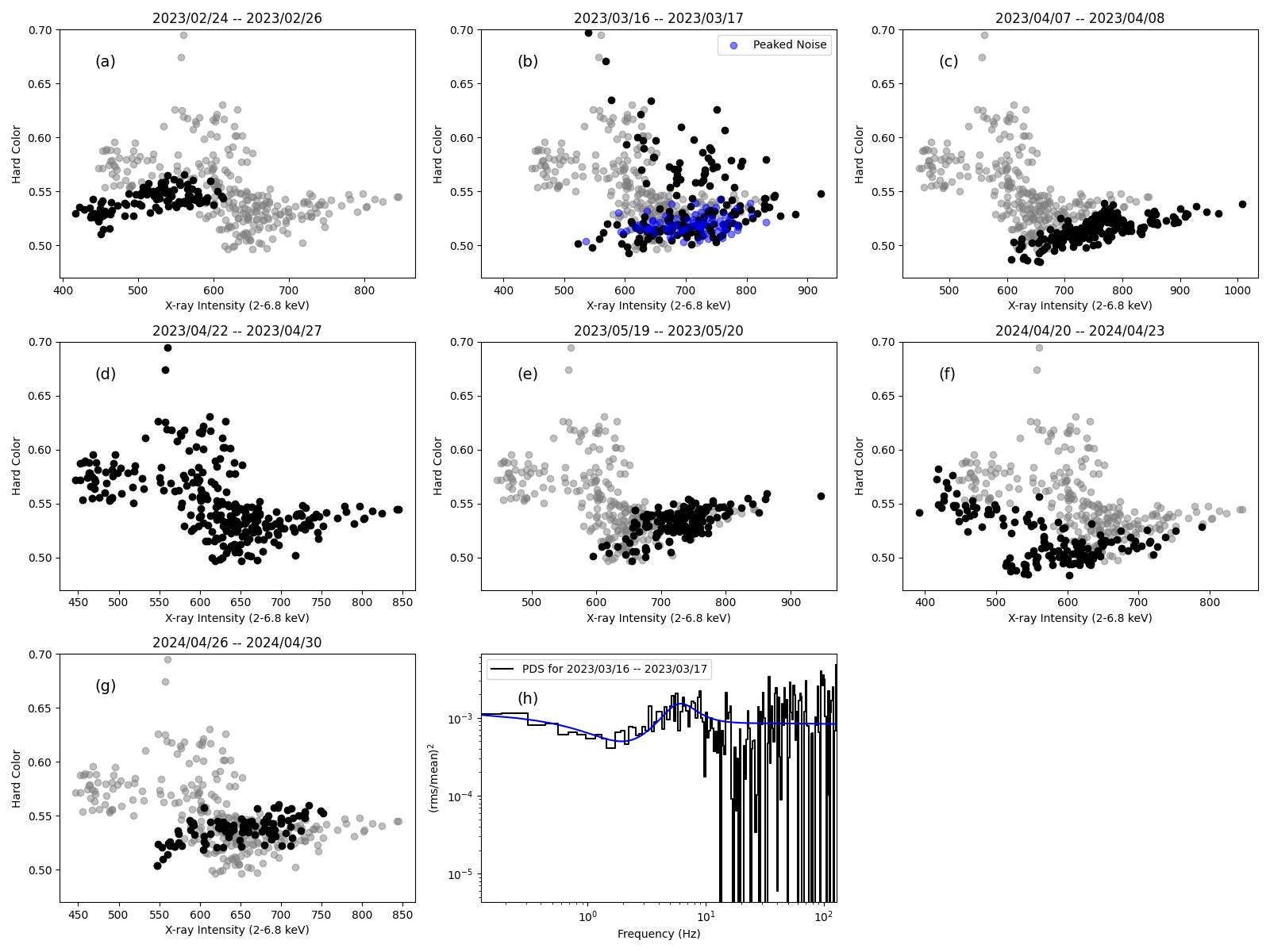}
    \caption{HIDs of all the observation sets over-plotted on the full track from the one observation set shown in panel d. Panel h shows the PDS with the peaked noise component seen in the highlighted portion of panel b.}
    \label{fig1}
\end{figure}   

\newpage
\section{Results}
The HIDs for each observation were visually inspected and sets of consecutive observations were plotted together such that any secular evolution in the HID could be observed (see Figure 1). Most sets of observations traced out portions of the HID, making it difficult to identify the state of the source. However, the observation set spanning 2023-04-22 to 2023-04-27 (panel d) traced out an entire Z-track making this the first time it has been observed in GX 13+1. Hence, the remaining observations were over-plotted on this complete track (grey markers) to identify the branches. The PDS of one of the observations (panel h) revealed a peaked noise component (PN). The very low and low frequency noise were modeled using a power-law and a zero-centered Lorentzian component, respectively, whereas the PN was modeled using a Lorentzian component. The resulting parameters showed that the PN has a central frequency of 5.44 $\pm$ 1.00 Hz and an FWHM of 4.99 $\pm$ 1.00 Hz with a quality factor Q $\sim$ 1.09 and rms amplitude of $\sim$ 3.8 \%. 

\section{Discussion \& Conclusion}

\cite{fridriksson2015} showed that GX 13+1 exhibited HID/CCD behavior of both Cyg-like and Sco-like Z sources, but the track was incomplete leading to ambiguity in classification. In our study using NICER data, we detect the complete Z-track of the source with three distinct branches during one set of observations (see Figure 1, panel d). The remaining observations showed only incomplete tracks mostly with a single branch. However, the dataset  obtained from 2023-03-16 to 2023-03-17 (panel b) showed two distinct branches, which could either be the HB and NB or a shifted NB and FB. 
This observation showed signatures of a PN at $\sim$ 5.4 Hz. It is difficult to classify this feature based on its HID location, as it could either be from the NB or the FB. GX 13+1 has previously shown a similar feature (band limited noise; BLN) in RXTE data throughout the two-branched track \citep{schnerr2003}. But, this is the first time that this feature has been observed down to 0.5 keV in GX 13+1.


The highest BLN rms amplitude for GX 13+1 measured by \citealt{schnerr2003} was $\sim$ 4--6\%. BLN was found for low Sa values, where Sa indicates the position of the observation along the CCD with Sa values increasing from the NB/top of atoll track to FB/lower part of atoll track (Sa of 2 is the soft apex/vertex). As Sa increased, the BLN became undetectable. Upon placing the rms amplitude of our peaked noise component in their BLN amplitude vs Sa track (see their Figure 13), we can note that our PN should have Sa $<$ 2 which indicates its location to be in the NB. As per our HID, this could be the true if what we are seeing is the HB and NB tracks (panel b). 

The shape of the Z-track we obtained shows a positive slope of the HB. Similar behavior of HB was seen in GX 340+0 \citep{pahari2024} using NICER data and they suggested that this could be either because of a lower contribution of the components that are active in the 6--20 keV energy band in the softer energy bands of NICER or the strong intrinsic absorption of the source.
This could also hold true in the case of GX 13+1 which has a high absorption column density ($>10^{22}$ cm$^{-2}$: \citealt{giridharan2024, saavedra2023}). The positive slope of the HB seen here could be due to the stronger presence of softer components in the selected energy domain or the high absorption column. Additional observations with longer exposure times are required to track the secular evolution of the Z-track of GX 13+1. 

\section{Acknowledgements}
This research has made use of data and/or software provided by the High Energy Astrophysics Science Archive Research Center (HEASARC), which is a service of the Astrophysics Science Division at NASA/GSFC.

\end{document}